\documentclass[12pt]{article}
\usepackage[cp1251]{inputenc}
\usepackage[russian]{babel}
\begin{document}
\large
\begin{center}
{\bf Expressions for Neutrino Wave Functions and Transition
Probabilities at Three Neutrino Oscillations in Vacuum and Some of
Their Applications}
\par
\vspace{0.3cm} Beshtoev Kh. M.
\par
\vspace{0.3cm} Joint Institute for Nuclear Research., Joliot Curie
6, 141980 Dubna, Moscow region and Institute of Applied
Mathematics and Automation  KBSRC of RAS, Nalchik,  Russia;
\end{center}
\vspace{0.3cm}

\par
Abstract

\par
We have considered three neutrino transitions and oscillations in
the general case and obtained expressions for neutrino wave
functions in three cases: with $CP$ violation, without $CP$
violations and the case when $\nu_e \leftrightarrow \nu_\tau$
transitions are absent (some works indicate on this possibility).
Then using the existing experimental data some analysis has been
fulfilled. This analysis definitely has shown that transitions
$\nu_e \leftrightarrow \nu_\tau$ cannot be closed for the Solar
neutrinos. However, this possibility may be realized by using the
mechanism of resonance enhancement neutrino oscillations in matter
(the Sun). But this possibility is not confirmed by the Solar
neutrinos spectrum (the Solar neutrinos spectrum is not distorted)
and the Day-Night effect (this effect is not observed). It was
found out that the probability of $\nu_e \leftrightarrow \nu_e$
neutrino transitions is positive defined value only if the angle
of $\nu_e, \nu_\tau$ mixing $\beta \le 15^o \div 17^o$.

%1.
\section{Introduction}

\par
The suggestion that, in analogy with $K^{o},\bar K^{o}$
oscillations, there could be neutrino-antineutrino oscillations (
$\nu \rightarrow \bar \nu$) was considered by Pontecorvo [1] in
1957. It was subsequently considered by Maki et al. [2] and
Pontecorvo [3] that there could be mixings (and oscillations) of
neutrinos of different flavors (i.e., $\nu _{e} \rightarrow \nu
_{\mu }$ transitions).
\par
In the general case there can be two schemes (types) of neutrino
mixings (oscillations): mass mixing schemes and charge  mixings
scheme (as it takes place in the vector dominance model or vector
boson mixings in the standard model of electroweak interactions)
[4].
\par
In the Standard theory of neutrino oscillations [5] it is supposed
that physically observed neutrino states $\nu_{e}, \nu_{\mu },
\nu_{\tau}$ have no definite masses and that they are directly
created as mixture of the $\nu_{1}, \nu_{2}, \nu_{3}$ neutrino
states. However the computation has shown that these neutrinos
have definite masses [4]. Then neutrino mixings are determined by
the neutrino mass matrix and neutrino mixing parameters are
expressed through elements of the neutrino mass matrix.
\par
In the scheme of charge mixings the oscillation parameters are
expressed through weak interaction couple constants (charges)and
neutrino masses [4].
\par
In the both cases the neutrino mixing matrix $V$ can be given [4]
in the following convenient form proposed by Maiani [6]:
\par
$$
{V {=} \left( \begin{array} {ccc}1& 0 & 0 \\
0 & c_{\gamma} & s_{\gamma} \\ 0 & -s_{\gamma} & c_{\gamma} \\
\end{array} \right)\!\! \left( \begin{array}{ccc} c_{\beta} & 0 &
s_{\beta} \exp(-i\delta) \\ 0 & 1 & 0 \\ -s_{\beta} \exp(i\delta)
& 0 & c_{\beta} \end{array} \right)\!\! \left( \begin{array}{ccc}
c_{\theta} & s_{\theta} & 0 \\ -s_{\theta} & c_{\theta} & 0 \\ 0 &
0 & 1 \end{array}\right)} , \eqno(1)
$$
$$
c_{e \mu} = \cos {\theta } , \quad s_{e \mu} =\sin{\theta}, \quad
c^2_{e \mu} + s^2_{e \mu} = 1 ;
$$
$$
c_{e \tau} = \cos {\beta }, \quad s_{e \tau} =\sin{\beta}, \quad
c^2_{e \tau} + s^2_{e \tau} = 1 ; \eqno(2)
$$
$$
c_{\mu \tau} = \cos {\gamma} , \quad s_{\mu \tau} =\sin{\gamma},
\quad c^2_{\mu \tau} + s^2_{\mu \tau} = 1 ;
$$
$$
 \exp(i\delta) = \cos{\delta } + i \sin{\delta} .
$$
Now we will come to computation of neutrino wave functions
$\Psi_{\nu_e}, \Psi_{\nu_\mu}, \Psi_{\nu_\tau}$ and a probability
of transitions (oscillations) of these neutrinos.

%2.
\section{General Expressions for Neutrino Wave Functions and Probabilities in
Dependence on Time at Three Neutrino Transitions (Oscillations) in
Vacuum}

\par
Using the above matrix $V$, we can connect the wave functions of
physical neutrino states $\Psi_{\nu_e}, \Psi_{\nu_\mu},
\Psi_{\nu_\tau}$ with the wave functions of intermediate neutrino
states $\Psi_{\nu_1}, \Psi_{\nu_2}, \Psi_{\nu_3}$ and write down
it in a component-wise form [5]:
\par
$$
\Psi_{\nu_l} = \sum^{3}_{k=1}V^{*}_{\nu_l \nu_k} \Psi_{\nu_{k}},
$$
$$
\Psi_{\nu_k} = \sum^{3}_{k=l}V_{\nu_k \nu_l} \Psi_{\nu_{l}},
\qquad l = e, \mu, \tau , \qquad k = 1 \div 3 ,
\eqno(3)
$$
where $\Psi_{\nu_{k}}$ is a wave function of neutrino with
momentum $p$ and mass $m_{k}$. We suppose that neutrino mixings
(oscillations) are virtual if neutrinos have different masses (if
we suppose that these transitions are real, as it is supposed in
the standard theory of neutrino oscillations, then it is necessary
to accept that expression (3) is based on a supposition that
masses difference of $\nu_{k}$ neutrinos is so small that coherent
neutrino states are formed in the weak interactions (computation
has shown that this condition is not fulfilled, i.e. neutrino as
wave packet is unstable and decays)).
\par
$$
\Psi_{\nu_{k}}(t) = e^{-i E_k t} \Psi_{\nu_{k}}(0) . \eqno(4)
$$
\par
Then
$$
\Psi_{\nu _{l}(t)} =\sum^{3}_{k=1} e^{-i E_k t} V^{*}_{\nu_l
\nu_k} \Psi_{\nu_{k}}(0)  . \eqno(5)
$$
Using unitarity of matrix $V$ or expression (3) we can rewrite
expression (5) in the following form:
\par
$$
\Psi_{\nu _{l}}(t) = \sum^{}_{l'= e,\mu, \tau} \sum^{3}_{k=1}
V_{\nu_{l'} \nu_k} e^{-i E_k t} V^{*}_{\nu_l \nu_k}
\Psi_{\nu_{l'}(0)} , \eqno(6)
$$
and introducing symbol $b_{\nu_{l}\nu _{l'}}(t)$
$$
b_{\nu_{l}\nu _{l'}}(t) = \sum^{3}_{k=1} V_{\nu_{l'} \nu_k} e^{-i
E_k t} V^{*}_{\nu_l \nu_k} , \eqno(7)
$$
we obtain
$$
\Psi_{\nu_{l}}(t) = \sum^{}_{l'=e, \mu, \tau}
b_{\nu_{l}\nu_{l'}}(t) \Psi_{\nu_{l'}}(0) , \eqno(8)
$$
where $b_{\nu_{l} \nu_{l'}}(t)$-is the amplitude of transition
probability $\Psi_{\nu_{l}}  \rightarrow  \Psi_{\nu_{l'}}$.
\par
\noindent And the corresponding transition probability
$\Psi_{\nu_{l}} \rightarrow \Psi_{\nu_{l'}}$ is:
\par
$$
P_{\nu_{l}\nu_{l'}}(t) =\mid \sum^{3}_{k=1} V_{\nu_l' \nu_k} e^{-i
E_k t} V^{*}_{\nu_l \nu_k} \mid^{2} . \eqno(9)
$$
\noindent It is obvious that
\par
$$
\sum^{}_{l'=e, \mu, \tau} P_{\nu _{l'} \nu_{l}}(t) = 1 .
\eqno(10)
$$
%2.1.
\subsection{Expressions for Neutrino Wave Functions and Probability
of $\nu_e, \nu_\mu, \nu_\tau \to \nu_e, \nu_\mu, \nu_\tau$
Transitions (Oscillations) with $CP$ Violation in Vacuum}

The wave functions of $\nu_e, \nu_\mu, \nu_\tau \to \nu_e,
\nu_\mu, \nu_\tau$ transitions with $CP$ violation have the
following form:
\par
1. for the case of $\nu_e \to \nu_e, \nu_\mu, \nu_\tau$
transitions:
$$
\Psi_{\nu_e \to \nu_e, \nu_\mu, \nu_\tau} (t)=[cos^2(\beta) cos^2
(\theta) exp(-i E_1 t)+cos^2(\beta) sin^2 (\theta)
$$
$$
exp(-i E_2 t)+ sin^2 (\beta) exp(-i E_3 t)] \Psi_{\nu_e}(0) +
$$
$$
+[cos(\beta) cos(\theta) exp(-i E_1 t) (-cos(\gamma) sin(\theta)-
$$
$$
-sin(\beta) exp(-i \delta) sin(\gamma) cos(\theta))+
$$
$$
+cos(\beta) sin(\theta) exp(-i E_2 t) (cos(\gamma) cos(\theta)-
$$
$$
-sin(\beta) exp(-i \delta) sin(\gamma) sin(\theta))+
$$
$$
+sin(\beta) exp(-i \delta) exp(-i E_3 t) sin(\gamma) cos(\beta)]
\Psi_{\nu_\mu}(0)+
$$
$$
+[cos(\beta) cos(\theta) exp(-i E_1 t) (sin(\gamma) sin(\theta)-
$$
$$
-sin(\beta) exp(-i \delta) cos(\gamma) cos(\theta))+
$$
$$
+cos(\beta) sin(\theta) exp(-i E_2 t) (-sin(\gamma) cos(\theta)-
$$
$$
-sin(\beta) exp(-i \delta) cos(\gamma) sin(\theta))+ \eqno(11)
$$
$$
+sin(\beta) exp(-i \delta) exp(-i E_3 t) cos(\gamma) cos(\beta)]
\Psi_{\nu_\tau}(0) .
$$
\par
2. for the case of $\nu_\mu \to \nu_e, \nu_\mu, \nu_\tau$
transitions:
$$
\Psi_{\nu_\mu \to \nu_e, \nu_\mu, \nu_\tau} (t)=[(-sin(\gamma)
sin(\beta) exp(i \delta) cos(\theta)-
$$
$$
-cos(\gamma) sin(\theta)) exp(-i E_1 t) cos(\beta) cos(\theta)+
$$
$$
+(-sin(\gamma) sin(\beta) exp(i \delta) sin(\theta)+
$$
$$
+cos(\gamma) cos(\theta)) exp(-i E_2 t) cos(\beta) sin(\theta)+
$$
$$
+sin(\gamma) cos(\beta) exp(-i E_3 t) sin(\beta) exp(i \delta)]
\Psi_{\nu_e}(0)+
$$
$$
+[(-sin(\gamma) sin(\beta) exp(i \delta) cos(\theta)-
$$
$$
-cos(\gamma) sin(\theta)) exp(-i E_1 t) (-cos(\gamma) sin(\theta)-
$$
$$
-sin(\beta) exp(-i \delta) sin(\gamma) cos(\theta))+
$$
$$
+(-sin(\gamma) sin(\beta) exp(i \delta) sin(\theta)+
$$
$$
+cos(\gamma) cos(\theta)) exp(-i E_2 t) (cos(\gamma) cos(\theta)-
$$
$$
-sin(\beta) exp(-i \delta) sin(\gamma) sin(\theta))+
$$
$$
+sin^2 (\gamma) cos^2 (\beta) exp(-i E_3 t)] \Psi_{\nu_\mu}(0)+
$$
$$
+[(-sin(\gamma) sin(\beta) exp(i \delta) cos(\theta)-
$$
$$
-cos(\gamma) sin(\theta)) exp(-i E_1 t) (sin(\gamma) sin(\theta)-
$$
$$
-sin(\beta) exp(-i \delta) cos(\gamma) cos(\theta))+
$$
$$
+(-sin(\gamma) sin(\beta) exp(i \delta) sin(\theta)+
$$
$$
+cos(\gamma) cos(\theta)) exp(-i E_2 t) (-sin(\gamma)
cos(\theta)-\eqno(12)
$$
$$
-sin(\beta) exp(-i \delta) cos(\gamma) sin(\theta))+
$$
$$
+sin(\gamma) cos^2 (\beta) exp(-i E_3 t) cos(\gamma)]
\Psi_{\nu_\tau}(0) .
$$
\par
3. for the case of $\nu_\tau \to \nu_e, \nu_\mu, \nu_\tau$
transitions:
$$
\Psi_{\nu_\tau \to \nu_e, \nu_\mu, \nu_\tau} (t)=[(-cos(\gamma)
sin(\beta) exp(i \delta) cos(\theta)+sin(\gamma) sin(\theta))
$$
$$
exp(-i E_1 t) cos(\beta) cos(\theta)+ (-cos(\gamma) sin(\beta)
exp(i \delta) sin(\theta)-
$$
$$
-sin(\gamma) cos(\theta)) exp(-i E_2 t) cos(\beta) sin(\theta)+
$$
$$
+cos(\gamma) cos(\beta) exp(-i E_3 t) sin(\beta) exp(i \delta)]
\Psi_{\nu_e}(0)+
$$
$$
+[(-cos(\gamma) sin(\beta) exp(i \delta) cos(\theta)+
$$
$$
+sin(\gamma) sin(\theta)) exp(-i E_1 t) (-cos(\gamma) sin(\theta)-
$$
$$
-sin(\beta) exp(-i \delta) sin(\gamma) cos(\theta))+
$$
$$
+(-cos(\gamma) sin(\beta) exp(i \delta) sin(\theta)-
$$
$$
-sin(\gamma) cos(\theta)) exp(-i E_2 t) (cos(\gamma) cos(\theta)-
$$
$$
-sin(\beta) exp(-i \delta) sin(\gamma) sin(\theta))+
$$
$$
+sin(\gamma) cos^2 (\beta) exp(-i E_3 t) cos(\gamma)]
\Psi_{\nu_\mu}(0)+
$$
$$
+[(-cos(\gamma) sin(\beta) exp(i \delta) cos(\theta)+
$$
$$
+sin(\gamma) sin(\theta)) exp(-i E_1 t) (sin(\gamma) sin(\theta)-
$$
$$
-sin(\beta) exp(-i \delta) cos(\gamma) cos(\theta))+
$$
$$
+(-cos(\gamma) sin(\beta) exp(i \delta) sin(\theta)-
$$
$$
-sin(\gamma) cos(\theta)) exp(-i E_2 t) (-sin(\gamma) cos(\theta)-
\eqno(13)
$$
$$
-sin(\beta) exp(-i \delta) cos(\gamma) sin(\theta))+
$$
$$
+cos^2 (\gamma) cos^2 (\beta) exp(-i E_3 t)] \Psi_{\nu_\tau}(0) .
$$
Now consider the case when the $CP$ violation is absent.  \\

%2.2.
\subsection{Expressions for Neutrino Wave Functions and Probability
of $\nu_e, \nu_\mu, \nu_\tau \to \nu_e, \nu_\mu, \nu_\tau$
Transitions (Oscillations) without $CP$ Violation in Vacuum}

If we do not take $CP$ violation into account, then the expression
for the amplitude of neutrino transitions has the following forms:
\par
1. If primary neutrinos are $\nu_e$ neutrinos, then for this
neutrino wave function for $\nu_e \to \nu_e$, $\nu_e \to \nu_\mu$,
and $\nu_e \to \nu_\tau$ transitions we get
$$
\Psi_{\nu_e \to \nu_e, \nu_\mu, \nu_\tau} (t) = [cos^2 (\beta)
cos^2 (\theta) exp(-i E_1 t) +
$$
$$
+cos^2 (\beta) sin^2 (\theta) exp(-i E_2 t) +
$$
$$
+sin^2 (\beta) exp(-i E_3 t) ] \Psi_{\nu_e}(0) + \eqno(14)
$$
$$
+[cos(\beta) cos(\theta) exp(-i E_1 t)  (-sin(\gamma) sin(\beta)
cos(\theta)-
$$
$$
-cos(\gamma) sin(\theta))+cos(\beta) sin(\theta) exp(-i E_2 t)
(-sin(\gamma) sin(\beta) sin(\theta)+
$$
$$
+cos(\gamma) cos(\theta))+sin(\beta) exp(-i E_3 t)  sin(\gamma)
cos(\beta)] \Psi_{\nu_\mu}(0)
$$
$$
+[cos(\beta) cos(\theta) exp(-i E_1 t)  (-cos(\gamma) sin(\beta)
cos(\theta)+sin(\gamma) sin(\theta))+
$$
$$
+cos(\beta) sin(\theta) exp(-i E_2 t)  (-cos(\gamma) sin(\beta)
sin(\theta)-sin(\gamma) cos(\theta))+
$$
$$
+sin(\beta) exp(-i E_3 t)  cos(\gamma)
cos(\beta)]\Psi_{\nu_\tau}(0) .
$$
\par
2. For the case of $\nu_\mu \to \nu_e, \nu_\mu, \nu_\tau$
transitions we get
$$
\Psi_{\nu_\mu \to \nu_e, \nu_\mu, \nu_\tau} (t) = [cos(\beta)
cos(\theta) exp(-i E_1 t)
$$
$$
(-sin(\gamma) sin(\beta) cos(\theta)-cos(\gamma) sin(\theta))+
$$
$$
+cos(\beta) sin(\theta) exp(-i E_2 t)  (-sin(\gamma) sin(\beta)
sin(\theta)+cos(\gamma) cos(\theta))+
$$
$$
+sin(\beta) exp(-i E_3 t)  sin(\gamma) cos(\beta)] \Psi_{\nu_e}(0)
+
$$
$$
+[(-sin(\gamma) sin(\beta) cos(\theta)-cos(\gamma) sin^2(\theta))
exp(-i E_1 t) +
$$
$$
+(-sin(\gamma) sin(\beta) sin(\theta)+cos(\gamma) cos^2(\theta))
exp(-i E_2 t) +
$$
$$
+sin^2 (\gamma) cos^2 (\beta) exp(-i E_3 t) ] \Psi_{\nu_\mu}(0)+
$$
$$
[(-sin(\gamma) sin(\beta) cos(\theta)- cos(\gamma) sin(\theta))
exp(-i E_1 t)
$$
$$
(-cos(\gamma) sin(\beta) cos(\theta)+sin(\gamma) sin(\theta))+
$$
$$
+(-sin(\gamma) sin(\beta) sin(\theta)+cos(\gamma) cos(\theta))
exp(-i E_2 t)   \eqno(15)
$$
$$
(-cos(\gamma) sin(\beta) sin(\theta)- sin(\gamma)
cos(\theta))+sin(\gamma) cos^2 (\beta)
$$
$$
exp(-i E_3 t) cos(\gamma)] \Psi_{\nu_\tau}(0) .
$$
\par
3. For the case of $\nu_\tau \to \nu_e, \nu_\mu, \nu_\tau$
transitions we get
$$
\Psi_{\nu_\tau \to \nu_e, \nu_\mu, \nu_\tau} (t) = [cos(\beta)
cos(\theta) exp(-i E_1 t)
$$
$$
(-cos(\gamma) sin(\beta) cos(\theta)+sin(\gamma) sin(\theta))+
cos(\beta) sin(\theta) exp(-i E_2 t)
$$
$$
(-cos(\gamma) sin(\beta) sin(\theta)-sin(\gamma) cos(\theta))+
$$
$$
+sin(\beta) exp(-i E_3 t) cos(\gamma) cos(\beta)] \Psi_{\nu_e}(0)
+
$$
$$
+ [(-sin(\gamma) sin(\beta) cos(\theta)-cos(\gamma) sin(\theta))
exp(-i E_1 t)
$$
$$
(-cos(\gamma) sin(\beta) cos(\theta)+ sin(\gamma)
sin(\theta))+(-sin(\gamma) sin(\beta) sin(\theta)+
$$
$$
+cos(\gamma) cos(\theta)) exp(-i E_2 t) (-cos(\gamma) sin(\beta)
sin(\theta)-sin(\gamma) cos(\theta))+
$$
$$
+sin(\gamma) cos^2 (\beta) exp(-i E_3 t) cos(\gamma)]
\Psi_{\nu_\mu}(0)+  \eqno(16)
$$
$$
+ [(-cos(\gamma) sin(\beta) cos(\theta)+sin(\gamma) sin(\theta))^2
exp(-i E_1 t)+
$$
$$
+(-cos(\gamma) sin(\beta) sin(\theta)-sin(\gamma) cos(\theta))^2
exp(-i E_2 t)+
$$
$$
+cos^2 (\gamma) cos^2 (\beta) exp(-i E_3 t)] \Psi_{\nu_\tau}(0) .
$$
Probability of $\nu_e \to \nu_e$ neutrino transitions obtained
from exp. (14) is given by the following expression:
$$
P_{\nu_e \to \nu_e} (t)= 1 - cos^4(\beta)sin^2(2 \theta) sin^2(- t
(E_1-E_2)/2) -
$$
$$
cos^2(\theta) sin^2(2 \beta) sin^2(- t (E_1-E_3)/2) - \eqno(17)
$$
$$
-sin^2(\theta) sin^2(2 \beta) sin^2(- t (E_2-E_3)/2) .
$$

Probability of $\nu_e \to \nu_\mu$ neutrino transitions obtained
from exp. (14) is given by the following expression:
$$
P_{\nu_e \to \nu_\mu} (t)=4 cos^2(\beta) cos(\theta) sin(\theta)
[-sin(\gamma) sin(\beta) sin(\theta)+cos(\gamma) cos(\theta)]
$$
$$
[sin(\gamma) sin(\beta) cos(\theta)+cos(\gamma) sin(\theta)]
sin^2(- t (E_1-E_2)/2)-   \eqno(18)
$$
$$
+4 cos^2(\beta) sin(\beta) cos(\theta) sin(\gamma) [sin(\gamma)
sin(\beta) cos(\theta)+cos(\gamma) sin(\theta)]
$$
$$
\cdot sin^2(- t(E_1-E_3)/2)-4 cos^2(\beta) sin(\beta) sin(\theta)
sin(\gamma) [-sin(\gamma) sin(\beta) sin(\theta)+
$$
$$
+cos(\gamma) cos(\theta)] sin^2(- t(E_2-E_3)/2) .
$$

Probability of $\nu_e \to \nu_\tau$ neutrino transitions obtained
from exp. (14) is given by the following expression:
$$
P_{\nu_e \to \nu_\tau} (t)=4 cos^2(\beta) cos(\theta) sin(\theta)
[-cos(\gamma) sin(\beta) cos(\theta)+ \eqno(19)
$$
$$
+sin(\gamma) sin(\theta)] [cos(\gamma) sin(\beta)
sin(\theta)+sin(\gamma) cos(\theta)] sin^2(- t (E_1-E_2)/2)-
$$
$$
-4 cos^2(\beta) cos(\theta) sin(\beta) cos(\gamma) [-cos(\gamma)
sin(\beta) cos(\theta)+sin(\gamma) sin(\theta)]
$$
$$
\cdot sin^2(- t (E_1-E_3)/2)+4 cos^2(\beta) sin(\theta) sin(\beta)
cos(\gamma) [cos(\gamma) sin(\beta) sin(\theta)+
$$
$$
+sin(\gamma) cos(\theta)] sin^2(- t (E_2-E_3)/2) .
$$
Now we consider of neutrino wave functions and a probability of
neutrino transitions at the absence of $\nu_e \to \nu_\tau$
transitions.

\subsection{Expressions for Neutrino Wave Functions and
Probability of $\nu_e, \nu_\mu, \nu_\tau \to \nu_e, \nu_\mu,
\nu_\tau$ Transitions (Oscillations) in Vacuum at the Absence of
$\nu_e \to \nu_\tau$ Transitions}

\par
If primary neutrinos are $\nu_e$ neutrinos and there are no
transitions between $\nu_e$ and $\nu_\tau$ neutrinos, i.e., these
transitions are closed, then there only $\nu_e \to \nu_e$, $\nu_e
\to \nu_\mu$, and $\nu_\mu \to \nu_\tau$ after $\nu_e \to \nu_\mu$
transitions can exist. Then the amplitude of these transitions has
the following form:
$$
\Psi_{\nu_e \to \nu_e, \nu_\mu, \nu_\tau} (t) = [cos^2 (\theta)
exp(-i E_1 t) + sin^2 (\theta) exp(-i E_2 t) ] \Psi_{\nu_e}(0) +
\eqno(20)
$$
$$
+[-cos(\theta) sin(\theta) cos(\gamma) exp(-i E_1 t) + cos(\theta)
sin(\theta) cos(\gamma) exp(-i E_1 t)] \Psi_{\nu_\mu}(0) +
$$
$$
cos(\theta) sin(\theta) sin(\gamma) [exp(-i E_1) - exp(-i E_2)]
\Psi_{\nu_\tau}(0)  .
$$
And probabilities of these neutrino transitions (oscillations) are
described by the following expressions (in reality after
transitions $\nu_e \to \nu_\mu$ there must be transitions between
$\nu_\mu \to \nu_\tau$):
\par
for $\nu_e \to \nu_e$:
$$
P(\nu_e \to \nu_e, t) = 1 - sin^2 (2\theta)[cos^2 (2\gamma) +
sin^2 (2\gamma)]sin^2 (L/L_{12}) .  \eqno(21)
$$
\par
for $\nu_e \to \nu_\mu$:
$$
P(\nu_e \to \nu_\mu, t) = 1 - sin^2(2\theta)cos^2 (2\gamma) sin^2
(L/L_{12}) . \eqno(22)
$$
\par
for $\nu_e \to \nu_\tau$:
$$
P(\nu_e \to \nu_\tau, t) = 1 - sin^2(2\theta) sin^2 (2\gamma)sin^2
(L/L_{12}) ,  \eqno(23)
$$
where
$$
L_{ik}(m) = 1.27 \frac{E_{\nu_e}(MeV)}{|m^2_i-m^2_k|(eV^2)} \qquad
L = c t , \eqno(24)
$$
$E_{\nu_e}$- is energy of primary neutrino and $E_k = \sqrt{m_k^2
+ p_{\nu_e}^2} \simeq p_{\nu_e} + \frac{m_k^2}{p_{\nu_e}}$, $i, k
= 1 \div 3$.

\par
If primary neutrinos are $\nu_\mu$ neutrinos and there are no
transitions between $\nu_e$ and $\nu_\tau$ neutrinos, then
$$
\Psi_{\nu_\mu \to \nu_e, \nu_\mu, \nu_\tau}(t)=[-cos(\theta)
exp(-i t E_1 ) cos(\gamma) sin(\theta)+
$$
$$
+sin(\theta) exp(-i t E_2) cos(\gamma) cos(\theta)]
\Psi_{\nu_e}(0)+ [cos^2(\gamma) sin^2(\theta) exp(-i t E_1 )+
$$
$$
+cos^2(\gamma) cos^2(\theta) exp(-i t E_2)+sin^2(\gamma) exp(-i t
E_3)] \Psi_{\nu_\mu}(0)+  \eqno(25)
$$
$$
[-cos(\gamma) sin^2(\theta) exp(-i t E_1 ) sin(\gamma)-cos(\gamma)
cos^2(\theta) exp(-i t E_2) sin(\gamma)+
$$
$$
sin(\gamma) exp(-i t E_3) cos(\gamma)] \Psi_{\nu_\tau}(0) .
$$

\par
If primary neutrinos are $\nu_\tau$ neutrinos, there are no
transitions between $\nu_e$ and $\nu_\tau$ neutrinos
$$
\Psi_{\nu_\tau \to \nu_e, \nu_\mu, \nu_\tau} (t)=[cos(\theta)
exp(-i t E_1 ) sin(\gamma) sin(\theta)-
$$
$$
-sin(\theta) exp(-i t E_2) sin(\gamma) cos(\theta)]
\Psi_{\nu_e}(0)+
$$
$$
+[-cos(\gamma) sin^2(\theta) exp(-i t E_1 ) sin(\gamma)- \eqno(26)
$$
$$
-cos(\gamma) cos^2(\theta) exp(-i t E_2) sin(\gamma)+ sin(\gamma)
exp(-i t E_3) cos(\gamma)] \Psi_{\nu_\mu}(0)+
$$
$$
[sin^2(\gamma) sin^2(\theta) exp(-i t E_1 )+
$$
$$
sin^2(\gamma) cos^2(\theta) exp(-i t E_2)+cos^2(\gamma) exp(-i t
E_3)] \Psi_{\nu_\tau}(0) .
$$

\section{Some Analysis of Neutrino Oscillation Possibilities}

\par
The value of the Solar neutrinos flow measured (through elastic
scattering) on SNO [7] is in a good agreement with the same value
measured in Super-Kamiokande [8].
\par
Ratio of $\nu_e$ flow measured on SNO (CC) to the same flow
computed in the frame work of SSM [9] ($E_\nu > 6.0 MeV$) is:
$$
\frac{\phi_{SNO}^{CC}}{\phi_{SSM2000}} = 0.35 \pm 0.02 . \eqno(27)
$$
This value is in a good agreement with the same value of $\nu_e$
relative neutrinos flow measured on Homestake (CC) [10] for energy
threshold $E_\nu = 0,814 MeV$.
$$
 \frac{\Phi^{exp}}{\Phi^{SSM2000}} = 0.34 \pm 0.03 .
\eqno(28)
$$
From these data we can come to a conclusion that the angle mixing
for the Sun $\nu_e$ neutrinos does not depend on neutrino energy
thresholds (0.8 $\div$ 15 MeV). Now it is necessary to know the
value of this angle mixing $\theta_{\nu_e \nu_\mu}$. Estimation of
the value of this angle can be extracted from KamLAND [11] data
and it is:
$$
sin^2 \theta_{\nu_e \nu_\mu} \cong 1.0 , \quad \theta  \cong
\frac{\pi}{4}, \quad \Delta m^2_{1 2} = 6.9 \cdot 10^{-5} eV,
\eqno(29)
$$
or
$$
sin^2 \theta_{\nu_e \nu_\mu} \cong 0.83 , \quad \theta = 32^o ,
\quad \Delta m^2_{1 2} = 8.3 \cdot 10^{-5} eV.
$$
The angle mixing for vacuum $\nu_\mu \to \nu_\tau$ transitions
obtained on Super-Kamiokande [12] for atmospheric neutrinos is:
$$
sin^2 2\gamma_{\nu_\mu \nu_\tau}  \cong  1, \quad \gamma \cong
\frac{\pi}{4} \quad \Delta m^2_{2 3} = 2.1 \div 2.5 \cdot 10^{-3}
. \eqno(30)
$$
Now we can estimate the third angle mixing for ${\nu_e \to
\nu_\tau}$ transitions by using exp. (17). For this aim we average
the time dependence of exp. (17) taking into account that the
Earth is moving over the elliptic orbit and then ($\bar {sin}^2(-
t (E_i-E_j)/2) = 1/2, \quad i,j = 1,2,3$)
$$
 \bar P_{\nu_e \to \nu_e} = 1 - \frac{1}{2}[cos^4(\beta)sin^2(2 \theta)
+ cos^2(\theta) sin^2(2 \beta) + sin^2(\theta) sin^2(2 \beta)] =
$$
$$
=1 - \frac{1}{2}[cos^4(\beta)sin^2(2 \theta) + sin^2(2 \beta)] .
 \eqno(31)
$$
\par
By substituting the value of $sin^2 (2\theta)$ in (31) from exp.
(29) and the value of $\bar P_{\nu_e \to \nu_e}$ from exps. (27),
(28) we get
$$
 1.30 \simeq  [cos^4(\beta) + sin^2(2 \beta)] .
 \eqno(32)
$$
From the above expression we can come to conclusion that
$$
\beta \leq \pi/4 , \eqno(33)
$$
i.e., this angle is close to the maximal angle $\pi/4$.
\par
If we suppose that $\nu_e \to \nu_\tau$ transitions are closed,
then we can use exp. (21) to estimate angle $\beta$. For this aim
we average the time dependence of exp. (21) taking into account
that the Earth is moving over the elliptic orbit, then
$$
\bar P(\nu_e \to \nu_e, t) = 1 - \frac{1}{2} sin^2 (2\theta) .
\eqno(34)
$$
\par
By substituting the value of $sin^2 (2\theta)$ from exp. (29) in
(34) we get
$$
\bar P(\nu_e \to \nu_e, t) \simeq  \frac{1}{2}  , \eqno(35)
$$
we come to a contradiction with exp (27), (28), i.e. with
experiments, but this contradiction can be removed by using the
mechanism of resonance enhancement of neutrino oscillations in
matter [8]. But this possibility is not confirmed by the Solar
neutrinos spectrum (the Solar neutrinos spectrum is not distorted)
and Day-Night effect (this effect is not observed). Unlikely it is
possible to obtain a flat neutrinos energy spectrum without
distortion in broad energy region $E = 1 \div 15 MeV$ by using
this mechanism.
\par
It is also detected that using exp. (17) we can obtain limitation
on value of angle $\beta$. For this purpose we have fulfilled
graphical modelling of this function by using the following values
for [11] $\theta = 32.45^o, \Delta m^2_{1 2}$, and [12] $ \Delta
m^2_{2 3}$ from exprs. (29), (30) for different values of $\beta =
10^o \div 45^o$ and $\Delta m^2_{13} = 10^{-5} eV^2, \quad
5.7\cdot 10^{-5} eV^2,\quad 8.3\cdot 10^{-4} eV^2$ at the average
electron neutrino energy $\bar E_{\nu_e} = 7 MeV$ and established
that the value for $P(\nu_e \to \nu_e, t)$ become a positive
defined value at $\beta = 15^o \div 17^o$ ($P(\nu_e \to \nu_e, t)
\simeq 0$ at some values of $t$). If $\beta \ge 15^o \div 17^o$,
then $P(\nu_e \to \nu_e, t)$ becomes negative at some values of
$t$. Since the value for the probability of $\nu_e \leftrightarrow
\nu_e$ transitions $P(\nu_e \to \nu_e, t)$ must be positive
defined one then, if in reality neutrino oscillations take place,
the value for $\beta$ must be $\beta \le 15^o \div 17^o$.

\section{Conclusion}

\par
We have considered three neutrino transitions and oscillations in
the general case and obtained expressions for neutrino wave
functions in three cases: with $CP$ violation, without $CP$
violations and the case when $\nu_e \leftrightarrow \nu_\tau$
transitions are absent (some works indicate on this possibility).
Then using the existing experimental data some analysis has been
fulfilled. This analysis definitely has shown that transitions
$\nu_e \leftrightarrow \nu_\tau$ cannot be closed for the Solar
neutrinos. However, this possibility may be realized by using the
mechanism of resonance enhancement neutrino oscillations in matter
(the Sun). But this possibility is not confirmed by the Solar
neutrinos spectrum (the Solar neutrinos spectrum is not distorted)
and the Day-Night effect (this effect is not observed). It was
found out that the probability of $\nu_e \leftrightarrow \nu_e$
neutrino transitions is positive defined value only if the angle
of $\nu_e, \nu_\tau$ mixing $\beta \le 15^o \div 17^o$.  \\

\par
{\bf References}\\

\par
\noindent 1. Pontecorvo B. M., Soviet Journ. JETP, 1957, v. 33,
p.549;
\par
JETP, 1958,  v.34, p.247.
\par
\noindent 2. Maki Z. et al., Prog.Theor. Phys., 1962, vol.28,
p.870.
\par
\noindent 3. Pontecorvo B. M., Soviet Journ. JETP, 1967, v. 53,
p.1717.
\par
\noindent 4. Beshtoev Kh. M., JINR Communication E2-2004-58,
Dubna, 2004;
\par
hep-ph/0406124, 2004.
\par
\noindent 5. Bilenky S.M., Pontecorvo B.M., Phys. Rep.,
C41(1978)225;
\par
Boehm F., Vogel P., Physics of Massive Neutrinos: Cambridge
\par
Univ. Press, 1987, p.27, p.121;
\par
Bilenky S.M., Petcov S.T., Rev. of Mod.  Phys., 1977, v.59,
\par
p.631.
\par
 Gribov V., Pontecorvo B.M., Phys. Lett. B, 1969, vol.28,
p.493.
\par
\noindent 6.  Maiani L., Proc. Intern. Symp.  on Lepton--Photon
Interaction,
\par
DESY, Hamburg. P.867.

\par
\noindent 7. Ahmad Q. R. et al., Internet Pub. nucl-ex/0106015,
June 2001.
\par
Ahmad  Q. R. et al., Phys. Rev. Lett. 2002, v. 89, p.011301-1;
\par
Phys. Rev. Lett.  2002,v.  89, p.011302-1.
\par
\noindent 8. Fukuda S. et al., Phys Rev. Lett., 2001, v.86,
p.5651;
\par
Phys. Lett.B, 2002, v.539 p.179.
\par
Koshio Y. (Super-Kamiokande Collab.), Proc. of 28-th Intern.
\par
Cosmic Ray Conf., Japan, 2003, v.1, p.1225.
\par
\noindent 9. Bahcall D. et al., The Astrophysical Jour. 2001,
v.555, p.990.
\par
\noindent 10. Davis R., Prog. Part. Nucl. Phys., 1994, v.32, p.13
\par
\noindent 11. Eguchi K. et al., Phys. Rev. Let. 2003, v.90,
021802;
\par
Mitsui T., 28-th Intern. Cosmic Ray Conf., Japan, 2003, v.1
p.1221.

\par
\noindent 12. Habig A., Proceedings of Inter. Cosmic Ray Conf.,
Japan, 2003,
\par
v.1, p. 1255;
\par
Kearns Ed. Super-Kamiokande Collaboration, Report on Intern. Conf.
\par
Neutrino 2004, Paris, 2004.

\end{document}